\documentstyle[aps,epsf,epsfig]{revtex}

\newcommand{\bq}{\begin{equation}}
\newcommand{\eq}{\end{equation}}
\newcommand{\bqn}{\begin{eqnarray}}
\newcommand{\eqn}{\end{eqnarray}}
\newcommand{\nb}{\nonumber}
\newcommand{\lb}{\label}

\begin{document}
\title{Gravitational Collapse of a Circularly Symmetric Stiff Fluid with
Self-Similarity in $2+1$ Gravity }
\author{A. Y. Miguelote,  ${}^3$
N. A. Tomimura   $^{3}$ and
Anzhong Wang   ${}^{1, \; 2}$}
\address{
${}^{1}$ Physics Department, Zhejiang University of Technology,
Hong Zhou, China\\
${}^{2}$ CASPER, Physics Department, Baylor University,
Waco, TX76798-7316\\
${}^{3}$ Instituto de F\'{\i}sica, Universidade Federal Fluminense,
Av. Litor\^{a}nea s/n, Boa Viagem, 24210-340, Niter\'{o}i,
RJ, Brazil}

\date{\today}
\maketitle

\begin{abstract}

Linear perturbations of homothetic self-similar stiff fluid 
solutions, $S[n]$, with circular symmetry in $2+1$ gravity are 
studied. It is found that, except for those with $n = 1$ and 
$n = 3$, none of them is stable and all have more than one 
unstable mode. Hence, {\em none of these solutions can be 
critical}.

\end{abstract}

\vspace{.6cm}



\vspace{1.cm}

\renewcommand{\theequation}{1.\arabic{equation}}
\setcounter{equation}{0}

\section{Introduction} 

Studies of the non-linearity of the
Einstein field equations near the threshold of black hole formation 
reveal very rich phenomena \cite{Chop93}, which are quite similar 
to critical phenomena in statistical mechanics and quantum field 
theory \cite{Golden}. These phenomena were first discovered by Choptuik 
in a study of a collapsing massless scalar field, and since that time
they have been found to exist in various other matter fields \cite{Wang01}.  
  
In this paper, we study the gravitational collapse of a circularly
symmetric stiff fluid with homothetic self-similarity (HSS) 
in $2+1$ gravity. It is well known that a {\em time-like} massless 
scalar field is energetically equivalent to a stiff fluid \cite{stiff}. 
In fact, introducing the quantities
\bq
\lb{1.3}
 w_{\mu} \equiv \frac{\phi_{,\mu}}
{\sqrt{\phi_{,\alpha}\phi^{,\alpha}}}
\;\;\;
\rho = p =   \frac{1}{2}\phi_{,\alpha}\phi^{,\alpha},
\;\; \left(\phi_{,\alpha}\phi^{,\alpha} > 0\right)
\eq
we can write the energy-momentum tensor (EMT) as
\bq
\lb{1.4}
T_{\mu\nu}^{\phi} = \phi_{,\mu} \phi_{,\nu} 
- \frac{1}{2}  g_{\mu\nu} \phi_{,\alpha}\phi^{,\alpha}   
=  \rho \left(2w_{\mu} w_{\nu} -  g_{\mu\nu}\right),
\eq
where $\phi$ denotes the scalar field and $(\;)_{,\mu} \equiv
\partial (\;)/\partial x^{\mu}$.
It should be emphasized that {\em the above equivalence is valid only 
when the massless scalar field is time-like}. 

The main motivation for the present study comes from a recent investigation
of the critical collapse of a stiff fluid in $3+1$ gravity  \cite{Brad02},
in which it was shown numerically that  a
massless scalar field and a stiff fluid exhibit quite different critical
phenomena near the threshold of black hole formation, given the equivalence
of Eq. (\ref{1.3}). In particular, the critical solution
of the scalar field has discrete self-similarity (DSS), and
the corresponding exponent $\gamma$   is $\gamma \simeq
0.374$ \cite{Chop93}, while for a stiff fluid, the critical solution has 
 HSS, and the corresponding exponent is $\gamma 
\simeq 0.94$ \cite{NC00}. It should be noted that in the latter
case, the  HSS (critical) solution is not stable with respect to
kink perturbations \cite{Hara01}.

In this paper, we  study the same problem, but in $2+1$ gravity.  
In particular,  we  first review some properties of
the self-similar solutions in terms of a stiff fluid   particularly 
focusing   on their extension across the  self-similar horizon (SSH).
This is crucial when we consider the boundary conditions for their linear 
perturbations. As we  show, in some cases the extension for the
scalar field is analytical, and the only change   is that 
the scalar field becomes space-like in the extended region, while for 
a stiff fluid such an extension  leads to a negative energy density and 
space-like velocity. Thus, for the latter, we have 
to drop the requirement of analytical extension. Because the background
is not analytical, there is no reason to require the
perturbations to be analytical across the SSH.  As we   show, 
removing the requirement of analyticity results in a   
different spectrum for the perturbations. In particular, we   find that,
except for the solutions with $n = 1$ and $n = 3$, all 
the self-similar solutions for a stiff fluid are unstable and have more 
than one unstable mode. Hence, {\em none of the self-similar solutions is 
critical for the gravitational collapse of a stiff fluid}, because, 
by definition, a critical solution has {\em one and only one} unstable mode.

It should be noted that, because the problem has been studied extensively 
by several authors, it is difficult to aviod repeating material that has
previously been presented, but we try to keep such repetition to a minimum.
In Ref. 4) and 5),  only linear perturbations of the 
self-similar solutions that have analytical extensions across the SSH
were studied, although some of those results are valid even for solutions
that do not have analytical extension. In this paper,  we  study 
perturbations of all self-similar solutions, both those with and those
without analytical extension, in terms of a stiff fluid.

\vspace{.5cm}

\renewcommand{\theequation}{2.\arabic{equation}}
\setcounter{equation}{0}

\section{Spacetimes with Homothetic Self-Similarity} 

The self-similar solutions of a stiff fluid are given by \cite{HW02,yasuda}
 \bq
 \lb{2.11}
 ds^{2} =  \frac{\left[(-\bar{u})^{n} 
      + (-\bar{v})^{n}\right]^{4c^{2}}}
 {2^{4c^{2}-1}\left(1 - c^{2}\right)^{2}} d\bar{u} d\bar{v} -
\left[(-\bar{u})^{2n} - (-\bar{v})^{2n}\right]^{2}d\theta^{2},
 \eq
where
 \bq
 \lb{2.12}
 \rho =  \frac{\rho_{0} (\bar{u}\bar{v})^{n-1}}
 {\left[(-\bar{u})^{n} + (-\bar{v})^{n}\right]^{2(3n-1)/n}},
 \eq
where $\rho_{0} \equiv 2^{4c^{2}} c^{2}$ and $n \equiv
1/[2(1-c^{2})]$, with $n$ (or $c$) being a free parameter. 
The above solutions 
in general are valid only in the region satisfying   $\bar{u} \le 0, \;
\bar{v} \le 0$ and $\bar{v} \ge \bar{u}$ in the ($\bar{u},
\bar{v}$)-plane, which is referred to as Region $II$ and is 
shown in Fig. 1.   Region $I$, in which $\bar{u} \le 0,\; \bar{v} 
\ge 0,\; |\bar{u}| \ge \bar{v}$, is considered  an 
extended region. The extension is analytical 
only when $n$ is an integer. However, when $n$ is an even number,  
Eq. (\ref{2.12}) shows that the energy density $\rho$ becomes 
negative in Region $I$. This is not physically feasible. 
Given this situation, one may be tempted to consider the   
extension obtained by   replacing $-\bar{v}$ with $|\bar{v}|$. 
However, this extension is not
 analytical. In fact, {\em no analytical and physically
reasonable extension exists for the stiff fluid solutions of
Eq. (\ref{2.12}), except in the case that $n$ is an odd integer}.
This fact is crucial when we consider the boundary conditions at 
$\bar{v} = 0$ of linear perturbations. 

It should be noted that the surface $\bar{v} = 0$ is a 
sonic line of the stiff fluid. This surface is also called a self-similar horizon,
that is, a null surface of constant $z$, defined in Eq. (\ref{3.3a}) 
\cite{Brad02}. In addition, this SSH  is  degenerate, as shown explicitly in 
Refs. 5) and 10). This   differs from the corresponding 
$(3+1)$-dimensional case \cite{Brad02}, in which the apparent horizon
forms after the SSH. For this reason, the boundary conditions for the 
linear perturbations in these two cases are different. This, in 
turn, affects the spectra of the perturbations.

 \begin{figure}[htbp]
\begin{center}
\label{fig4}
\leavevmode
\epsfig{file=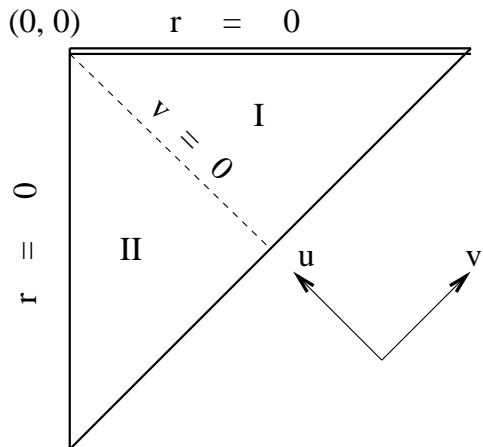,width=0.35\textwidth,angle=0}
\caption{The Penrose diagram for the solutions given  by
Eq. (\ref{2.12}). Region $I$ is the extended region.
When $n  = 2m + 1$, where $m$ is an integer, the dashed line 
$v = 0$ (or $\bar{v} = 0$) represents an apparent horizon, 
and the spacetime is singular on the
horizontal double line $r \equiv  (-\bar{u})^{2n} - 
(-\bar{v})^{2n} =0$. For $n \not= 2m + 1$, 
the extension is not analytical.}
\end{center}
\end{figure}

\vspace{.5cm}

\renewcommand{\theequation}{3.\arabic{equation}}
\setcounter{equation}{0}

\section{Linear Perturbations}

Let us first note that, when we consider
perturbations of the self-similar solutions of Eq. (\ref{2.12}), we
consider only Region $II$, where the equivalence (\ref{1.3}) 
between the stiff fluid and the massless scalar field
is always valid, because in this region 
the massless scalar field is always time-like. To aviod repetition
and also for the sake of simplicity, 
we treat only  a massless scalar field in this
section. The corresponding perturbations for the stiff fluid can
be easily obtained from the one-to-one correspondence given by
Eq. (\ref{1.3}). With our scope thus focused, we first introduce
 the null coordinates $u$ and $v$ via the relations
 \bq
 \lb{3.1}
 u = -  (-\bar{u})^{2n},\;\;\;
 v = -   (-\bar{v})^{2n}.
 \eq
Then, we find that the metric in Region $II$ for the background solutions
of Eq. (\ref{2.12}) in terms of $u$ and $v$ takes the   form $ds^{2} = 
2e^{2\sigma}du dv - r^{2}d\theta^{2}$, where
 \bqn
 \lb{3.3a}
 \sigma(u,v) &=& 2c^{2}ln\left[\frac{1}{2}
 \left(z^{1/4} + z^{-1/4}\right)\right],\nb\\
 r(u,v) &=& (-u)s(u,v),\;\;\;
s(u,v) = 1 - z, \;\;\; z \equiv \frac{v}{u} \ge 0,\\
 \lb{3.3b}
 \phi(u, v) &=& c \ln(-u) + \varphi(z), \;\;
 \varphi(z) = -2c\ln\left(1 + z^{1/2}\right).
 \eqn

Taking the above self-similar solutions as the background, let us
now     study   their linear perturbations  of the form
\bq
\lb{3.4}
F(z, \tau) = F_{0}(z) + \epsilon \, F_{1}(z)e^{k\tau}, 
\eq
where $F \equiv \{\sigma, \; s, \; \varphi\}$, $\epsilon$ is a very 
small real constant, and
$ \tau \equiv - \ln({-u}/{u_{0}})$,
with $u_{0}$ being   constant with the dimension of 
length. Without loss of generality, in the following we   set 
it to 1.
Quantities with the subscript ``1" represent perturbations, and those
with ``0" represent the self-similar solutions given by
Eqs. (\ref{3.3a}) and  (\ref{3.3b}).  Modes with $Re(k) > 0$ grow
with $\tau$ and are referred to as unstable
modes, while those with $Re(k) < 0$ decay and are referred to
as stable modes.   

Our treatment of the  perturbations is identical to that given in
Sec. 4.3 of Ref. 5), which is referred to as Paper I
hereafter. The only difference is in the boundary conditions on the
SSH,   $z = 0$. As shown in the last section,
the extension of the background solutions across this surface is
analytical only when $n$ is an odd integer. Thus, except in the case that
$n$ is an odd integer,  there is no reason to require that
the perturbations be analytical across $z = 0$. Therefore, to impose 
the boundary conditions properly, we
have to distinguish the two cases $n = 2m + 1$ and $n \not=
2m + 1 \; (m = 0, 1, 2, ...)$.

\vspace{.5cm}

\noindent{\bf Case III.A $\; n = 2m + 1$:}
When $n = 2m + 1$, the background solutions are analytical across
the hypersurface $z = 0$. Therefore, in this case, the boundary conditions are
the same as those  given by Eqs. (105) and Eq. (108) of Paper I.
Thus,  the analysis given in Sec. 4.3 of Paper I is
applicable to this case. In particular, it can be seen that, for
any given $m$, these background solutions have $N$ unstable modes,
where $N$ is given by
$N = n - 3 = 2(m -1)$.
Thus, the solutions with  $n = 1 $ and $n= 3 $ (or $m = 0, 1$) are
stable, while those with $n \ge 5$ (or $m \ge 2$) are not, 
and all of them have more than one unstable mode.
Hence, none of these solutions is  critical.

\vspace{.5cm}

\noindent{\bf Case III.B $\; n \not= 2m + 1$:}
When $n \not= 2m + 1$, the background solutions are not analytical
across the hypersurface $z = 0$, and there is no
reason   to require that the perturbations be analytical. Instead,
we  require that $s_{1},\; \sigma_{1}$ and $\varphi_{1}$ be
finite there. In addition, we stipulate that no matter emerge
from the self-similar horizon. Thus, we have the following conditions:  
\bq
\lb{3.7}
s_{1}(0),\; \sigma_{1}(0), \; {\mbox{and}} \; \varphi_{1}(0) \;\;
{\mbox{are  all   finite, and }} \;\;
 \lim_{z\rightarrow 0}{z^{\chi} \frac{d\varphi_{1}(z)}{dz}} \rightarrow 0,  
\eq
where $\chi \equiv c^{2}$. The boundary conditions at 
the symmetry axis $r = 0$ (or $z = 1$)  are 
the same as these given in Eq. (105) of Paper I.

Once the boundary conditions are determined, following Paper I we
can study the three cases (a) $k = 1$, (b) $k = l + 1/2$, and (c)
$k \not= 1, \; l + 1/2$ separately, where $l$ is a non-negative
integer.   The perturbations for Case (a) are given 
by Eqs. (110), (111), (114)-(118), while those for Cases (b) 
and (c) are given by Eqs. (120), (121) and (125) of Paper I.  
Perturbations due to the gauge transformations (79) are given by Eq. (109),
which can be obtained from Eqs. (120), (121) and (125) by setting
$c_{1} = 0 = c_{2}$, where $c_{1}$ and $c_{2}$ are  free parameters.
Thus, to exclude the gauge modes from the spectrum of perturbations, 
in the following we  restrict ourselves to the case in which 
\bq
\lb{cd4}
\left|c_{1}\right|^{2} + \left|c_{2}\right|^{2} \not= 0. 
\eq

The analyses of Cases 
(a) and (b) are similar to those given in Paper I. In particular,
 no perturbations in these two cases satisfy the boundary conditions.

In Case (c), where $k \not= 1, l + 1/2$, following the analysis given 
 between Eqs. (120) and (128) of Paper I, one can show that the 
boundary conditions given by Eq. (105) of Paper $I$ at the symmetry 
axis   require $c_{1} = 0$, 
where $c_{1}$ is an integration constant, given in Eq. (125).
Then, for $s_{1}(z)$ to be finite at   $z = 0$ it is necessary that  
$Re(k) < 1$,
as one can see from Eq. (120). On the other hand, it can
be shown that
\bqn
\lb{3.10}
\varphi_{1}(z) & \rightarrow & - \left(c_{2}A(k) - c_{0} c\right)
 z^{1/2-k},\nb\\
\frac{d\varphi_{1}(z)}{dz}
  & \rightarrow &   \frac{1}{2} \Big\{   \, cc_{0}\left[   z^{-1/2}
                    + 2(1-k)z^{-k} \right]
   + c_{2}(2k-1)A(k) z^{-1/2 - k}\nb\\
  &  &                   + \left[   cc_{0}(10k -7)
                                                  - c_{2}(1-k)A(k)
                                          \right] \, z^{1/2-k}
                                \Big\},\nb\\
\sigma_{1}(z)
   & \rightarrow &   \frac{\chi}{c} \left( A(k)c_{2} - c c_{0} \right) \,
                       z^{1/2 - k} \nb\\
& & 
                   + \frac{1}{2} \Big[   A(k)\chi(2k-1)c_{2}
                                       - (1-\chi -k)c_{0}
                                 \Big] \, z^{-k},
\eqn
as $z \rightarrow 0$,  where $A(k) \equiv
\Gamma(k-1/2)/[\Gamma(k)\Gamma(1/2)]$. Thus,   the conditions
(\ref{3.7}) require
\bq
\lb{4.37}
   Re(k) < \chi - \frac{1}{2},\; \chi > \frac{1}{2},\;
   Re(k) < \frac{1}{2}, \;
 c_{2} = \frac{1-\chi -k}{\chi(2k-1)A(k)}c_{0}.
\eq
Clearly, for any given $c$ (or $\chi \equiv c^{2}$), there exist
infinitely many modes with $Re(k) > 0$ that satisfy the
conditions  (\ref{cd4}) and (\ref{4.37}). Therefore, all of these 
solutions are unstable with respect to the linear perturbations. Since 
each of them has an infinite number of unstable modes, they too cannot 
be critical solutions.


\vspace{.5cm}

\section{Summary and Concluding Remarks}

In this paper, we have studied a collapsing stiff fluid
with HSS in ($2+1$)-dimensional circular spacetimes. 
The solutions are usually 
defined only in Region $II$ [cf. Fig. 1], and Region $I$ is an 
extended region. The extension is analytical only for some of 
the self-similar solutions. In particular, for the
solutions for which  $n$ is not   an integer,  it cannot
be analytical. 

In the case that $n$ is an even integer, i.e., $n = 2m$, 
the extension is analytical, but
the stiff fluid has negative energy density, and its velocity 
becomes space-like in the extended region. This is clearly physically unfeasible. 
Thus, in this case,  we must remove the requirement of analytical 
extension. This results in a  considerable difference from the case of a 
massless scalar field when we consider boundary conditions on 
the self-similar horizon for perturbations. Recall that, for the scalar
field, the extension is analytical, and the only change   
is that the scalar field becomes space-like in the extended 
region, which is physically feasible. 
Since the background solutions with $n = 2m$ for a 
stiff fluid are  not analytical across $z = 0$, there is no 
reason to require that their perturbations be analytical there, 
in contrast to those for the scalar field. Without requiring the 
analyticity of the perturbations  across the SSH, 
we showed that {\em the background solutions with $n = 2m $ for
a stiff fluid are unstable and have an infinite number of unstable 
modes}. However, in the case of a scalar field, the requirement 
of analytical perturbations leads to the result that solutions with $n = 2m$ 
have $N$ unstable modes, where $N \equiv n -3$. Thus, the solution 
with $n = 4$ has one and only one unstable mode, and hence by definition, 
it is a critical solution for the collapse of the scalar field 
\cite{HW02}. Contrastingly, all the solutions  with either $n \not= 2m$ or
$n \not= 2m + 1$ for both a stiff fluid and a scalar field are 
non-analytical across $z = 0$, as are their linear
perturbations. As a result, all of these solutions are unstable 
with respect to  linear perturbations and have an infinite number of unstable 
modes.   
When $n = 2m +1$, the extension of the background solutions for a 
stiff fluid is analytical across the SSH. In this
case, one may require that their linear perturbations   also be
analytical. By doing so, we found that for any given $m$, 
a self-similar solution  has $N$ unstable modes, where 
$N =  2(m -1)$. Thus, {\em the self-similar solutions with 
$m = 0$ and $m = 1$ are stable, while those with $m \ge 2$ are   
unstable, and all  have more than one unstable mode}.

In review of all the above findings, we can see that all the self-similar
solutions, except those with $n = 1$ and $n = 3$, of a stiff fluid 
given by Eq. (\ref{2.12}) are unstable
with respect to linear perturbations, and each of them has more than one
unstable mode. Since a critical
solution, by definition, has one and only one unstable mode, we
conclude that {\em none of these self-similar solutions can be
critical for the gravitational collapse of a stiff fluid with 
circular symmetry in $2+1$ gravity}. This contrasts  with the situation for
its  $3+1$ dimensional counterpart, where the critical solution
does have HSS \cite{Brad02,NC00}.

Therefore, although between a stiff fluid and a massless 
scalar field there exists an one-to-one correspondence,
namely Eq. (\ref{1.3}), this equivalence is only local,
and it is restricted to  region(s) in which the scalar field is
time-like.  Globally, the spacetimes for these two cases can
be quite different. In the case considered in this paper,
this results in different boundary conditions, due to their 
properties in the extended region across the SSH. It is exactly 
this difference that makes the spectra of their perturbations different. 
 
In the 4D case, similar results were found. In particular, 
in the collapse of a scalar field, the critical solution is periodic
and can be time-like only in some regions \cite{Chop93}. Thus, this
DSS critical solution is equivalent to a stiff fluid only in these
particular regions, and it is not a stiff fluid solution in the
whole spacetime. For this reason,  it cannot be critical for the 
gravitational collapse of a stiff fluid. Instead, it was shown 
that the critical solution for a stiff fluid has HSS \cite{Brad02},
subject to the instability with respect to kink perturbations
\cite{Hara01}.

Our results obtained in this paper also  clearly show that if 
critical phenomena are exhibited in  the gravitational collapse of a stiff 
fluid in $2+1$ gravity, the critical solution must  have either
discrete self-similarity or no self-similarity at all. 
Certainly, it is also possible that  critical 
collapse does not exist at all in the collapse of a stiff fluid. 
In any case, it would be very interesting to investigate 
these problems and obtain a definitive answer to the 
questions.

\section*{ Acknowledgments}  

The author (AW) would like to express
his gratitude for the financial support from Baylor University through the
2005 Summer Sabbatical. Part of the work was done when 
one (AW) of the authors was visiting   
Zhejiang University of Technology. He would like 
to express his gratitude to the University for hospitality.

\end{document}